\documentclass[
 reprint,
 amsmath,amssymb,
 superscriptaddress,
 aps,
 pra,
 nofootinbib,
]{revtex4-2}

\usepackage{graphicx}
\usepackage{dcolumn}
\usepackage{bm}

\usepackage{physics2}
\usephysicsmodule{ab}
\usephysicsmodule{op.legacy}
\usephysicsmodule{braket}
\usephysicsmodule{nabla.legacy}
\usepackage{derivative,fixdif}

\usepackage{tensor}
\usepackage{mathtools}

\usepackage{siunitx}
\usepackage{appendix}
\usepackage[colorlinks,linkcolor=blue,urlcolor=blue, citecolor=blue]{hyperref}
\usepackage[whole]{bxcjkjatype}
\usepackage[T1]{fontenc} 
\usepackage{comment}

\usepackage{mymcro}

\begin{document}

\title{A low-loss telecom-band nanofiber cavity for interfacing Yb atomic qubits}

\author{Seitaro Horikawa}
\affiliation{Nanofiber Quantum Technologies, Inc. (NanoQT), 1-22-3 Nishiwaseda, Shinjuku-ku, Tokyo 169-0051, Japan.}
\affiliation{Department of Applied Physics, Waseda University, 3-4-1 Okubo, Shinjuku-ku, Tokyo 169-8555, Japan.}
\author{Shinya Kato}
\altaffiliation[Present address: ]{RIKEN center for Quantum Computing (RQC), Wako, Saitama 351-0198, Japan.}
\affiliation{Nanofiber Quantum Technologies, Inc. (NanoQT), 1-22-3 Nishiwaseda, Shinjuku-ku, Tokyo 169-0051, Japan.}
\author{Ryotaro Inoue}
\affiliation{Nanofiber Quantum Technologies, Inc. (NanoQT), 1-22-3 Nishiwaseda, Shinjuku-ku, Tokyo 169-0051, Japan.}
\author{Takao Aoki}
\affiliation{Nanofiber Quantum Technologies, Inc. (NanoQT), 1-22-3 Nishiwaseda, Shinjuku-ku, Tokyo 169-0051, Japan.}
\affiliation{Department of Applied Physics, Waseda University, 3-4-1 Okubo, Shinjuku-ku, Tokyo 169-8555, Japan.}
\author{Akihisa Goban}
\email{akihisa.goban@nano-qt.com}
\affiliation{Nanofiber Quantum Technologies, Inc. (NanoQT), 1-22-3 Nishiwaseda, Shinjuku-ku, Tokyo 169-0051, Japan.}
\author{Hideki Konishi}
\email{hideki.konishi@nano-qt.com}
\affiliation{Nanofiber Quantum Technologies, Inc. (NanoQT), 1-22-3 Nishiwaseda, Shinjuku-ku, Tokyo 169-0051, Japan.}

\begin{abstract} 

We demonstrate the fabrication of an optical nanofiber cavity designed for efficient interface with ytterbium (Yb) atoms at telecom-wavelength transitions.
Replacing the conventional hydrogen-oxygen flame with a deuterium-oxygen flame in the heat-and-pull method suppresses hydroxyl-induced absorption losses and enables low-loss nanofiber production with minimal modifications to the existing fabrication system.
%By replacing the conventional hydrogen–oxygen flame with a deuterium–oxygen flame in the heat-and-pull method, we suppress hydroxyl (OH)-induced absorption losses, achieving low-loss nanofiber transmission with minimal modifications to an existing fabrication system.
%Low transmission loss through a nanofiber at these wavelengths is achieved using the heat-and-pull method with a deuterium-oxygen flame as a heat source, requiring only minimal modifications to existing fabrication setups. 
Using this technique, we fabricate a nanofiber cavity at 1389~nm that exhibits an intrinsic round-trip loss of $0.31(2)\%$ and a finesse of $2.0(1)\times 10^3$.%2005(148). 
This performance corresponds to a projected cooperativity of 90 when interfaced with Yb atoms, indicating that the cavity is well suited for efficient atom-photon coupling at telecom wavelength transitions.
Our results establish a practical route for developing fiber-integrated atom–photon interfaces in the telecom band, a critical step toward scalable quantum communication and distributed quantum computing.
%\textcolor{cyan}{[AG: plz add Kato san's current affiliation as a foot note: RIKEN center for Quantum Computing (RQC), Wako, Saitama 351-0198, Japan]}

\end{abstract}

\maketitle
%\tableofcontents

\section{Introduction}
\label{sec:intro}

Neutral-atom arrays have emerged as a promising platform for quantum information processing~\cite{Bluvstein2022-ob, Bluvstein2024-ml, Norcia2023-ut, Muniz2025-yu, Reichardt2024-uf, Graham2023-kq}. 
While recent developments demonstrate a rapid increase in the number of qubits exceeding 1000 in a single unit~\cite{Norcia2024-wf, Gyger2024-on, Manetsch2024-gs, Pichard2024-dv}, scaling it up to millions, a typical size required to perform fault-tolerant quantum computing, remains challenging due to technical limitations such as the laser power for optical tweezers. 
To this end, modular quantum computing, in which quantum processing units with a moderate number of qubits are connected through photonic links, has been proposed as a promising approach~\cite{Monroe2014-sy, Covey2023-yw, Sunami2025-on}. 
In this architecture, each unit incorporates an interconnect device which interfaces stationary memory qubits and photonic qubits for remote entanglement generation and must operate sufficiently fast so as not to become a bottleneck of the quantum computation. 

For efficient interconnection, cavity quantum electrodynamics (QED) systems that support strong atom-photon coupling and high photon collection efficiency are a suitable platform.
An optical nanofiber cavity, formed by two fiber Bragg gratings (FBGs) as end mirrors and a nanofiber region in between, is one of the ideal devices for such an interface. 
Strong coupling with single cesium atoms in a nanofiber cavity has been demonstrated~\cite{Kato2015-kj}. 
Compared to other cavity QED platforms such as free-space and on-chip waveguide cavities, nanofibers provide a uniquely long active area, enabling uniform optical coupling for a one-dimensional array of more than 100 atoms per cavity---an essential feature for scalable multiplexed operations \cite{Sunami2025-on}.
Finally, their excellent connectivity to standard optical fibers offers seamless integration to optical fiber-based quantum networks.

Among various atomic species, ytterbium (Yb) is considered to be a promising candidate as atomic qubits thanks to its robustness against noise and a rich energy structure, including metastable states~\cite{Jenkins2022-hn, Ma2022-zy, Okuno2022-ts}. In addition, Yb is also suitable for quantum networking as it possesses transitions with telecom-band wavelengths, in particular, 1389~nm, 1480~nm, and 1539~nm connecting (6s6p)${}^3$P$_J$ and (5d6s)${}^3$D$_J$ states~\cite{Covey2019-ol, Li2024-bu, Sunami2025-on}. By interfacing Yb atomic qubits with these wavelengths, one can directly leverage existing optical-fiber networks for long-distance quantum communications. Therefore, the development of optical nanofiber cavities at the telecom-band wavelengths is a key milestone to realize large-scale fault-tolerant quantum computing and quantum networking.

The key figure of merit of a cavity QED system is cooperativity $C$, which is proportional to $\sigma_0 A_\mathrm{eff}^{-1}\mathcal{F}$ with the optical cross section of the atom $\sigma_0$, the effective mode area $A_\mathrm{eff}$, and the cavity finesse $\mathcal{F}$. 
Given the transition strength of the relevant transitions of Yb~\cite{Sunami2025-on} and the geometrical design of our nanofiber cavity, %resulting in $\sigma_0/A_\mathrm{eff}\sim 1$, 
the cavity finesse needs to be $\mathcal{F}>1000$ to reach well into the strong coupling regime of $C\gg 1$. 
This requires a suppression of a round-trip cavity loss below 0.6\%. %\textcolor{cyan}{[AG: I made the logic of setting F=1000 target vague, since not directly tied to explicit cooperativity value, (F=1000--> C=45). Let me know if you have better way to phrase this.]}

Such performance is readily achievable at visible to infrared wavelengths using the state-of-the-art nanofiber fabrication process using a flame generated by a hydrogen (H$_2$) and oxygen (O$_2$) gas mixture~\cite{Ruddell2020-yn, Horikawa2024-ra}. %However, the H$_2$-O$_2$ flame significantly increases the hydroxyl (OH) content in the fiber, which has an absorption band around 1380~nm, which is an overtone of the fundamental vibrational mode at 2760~nm~\cite{Humbach1996-pb}. This prevents the fabrication of low-loss nanofiber cavities at the telecom-band wavelengths.
However, it is not suitable for fabricating nanofibers at the telecom-band wavelength relevant to the Yb transitions, due to an absorption band of hydroxyl (OH) content implanted in the fiber during processing.

In this work, we demonstrate the suppression of OH-induced losses by replacing the H$_2$-O$_2$ flame with a flame generated from deuterium (D$_2$) and O$_2$, requiring only minimal modifications to the conventional fabrication system. 
This enables the fabrication of low-loss,  telecom-band nanofiber cavities suitable for interfacing with Yb atomic qubits. 
This paper is structured as follows. 
In Section~\ref{sec:nanofiber}, we first present a proof of concept for the D$_2$-O$_2$ flame approach to suppress losses at the telecom-band wavelengths. 
In Section~\ref{sec:cavity}, we fabricate a nanofiber cavity at 1389~nm using this technique and demonstrate that it achieves a finesse high enough for an efficient atom-photon interface. 
Finally, in Section~\ref{sec:conclusion}, we conclude with a discussion of potential improvements and future directions.

\section{Heat and pull with a deuterium-oxygen flame}
\label{sec:nanofiber}

\begin{figure}[t]
    \includegraphics{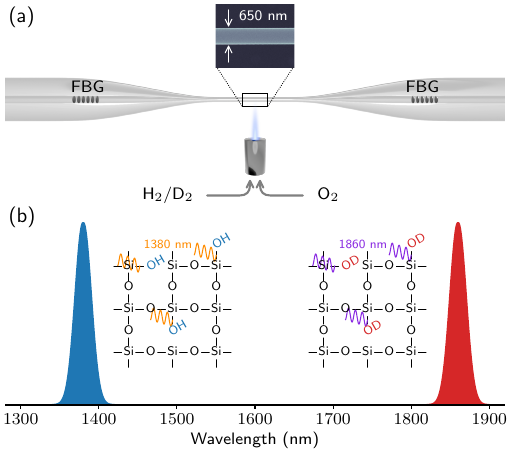}
    \centering
    \caption{(a)~An illustration of the heat-and-pull method for nanofiber fabrication. An optical fiber is stretched while the center part is heated with a flame. The flame is generated by the combustion of hydrogen or deuterium with oxygen (see main text). For the fabrication of nanofiber cavities, two FBGs are inscribed beforehand. The inset shows a scanning electron microscope image of a nanofiber.
    (b)~The 1st overtone absorptions of the OH (blue) and OD (red) groups. The insets represent the chemical structure of silica with OH and OD groups implanted. Si-OH and Si-OD bonds absorb light around 1380~nm and 1860~nm, respectively.}
    \label{fig:concept}
\end{figure}

\begin{figure}[t]
    \includegraphics{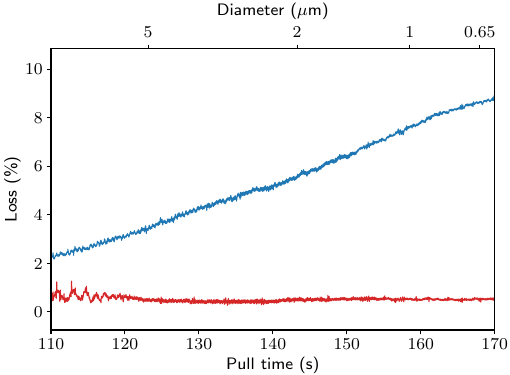}
    \centering
    \caption{Transmission losses at 1389~nm during nanofiber pulling. The blue and red traces correspond to samples pulled with an H$_2$-O$_2$ and D$_2$-O$_2$ flame, respectively.
    %(b)~Wavelength-dependent loss. The blue (red) points represent transmission losses measured at various wavelengths after 170~s of pulling with a H$_2$-O$_2$ (D$_2$-O$_2$) flame. The gray dashed line shows an absorption spectrum of Si-OH, reproduced from~\cite{Humbach1996-pb} and rescaled for clarity.
    }
    \label{fig:nanofiber}
\end{figure}

We fabricate nanofibers using the heat-and-pull method, where an optical fiber is pulled from both ends while the center is heated to soften the silica, as shown in Fig.~\ref{fig:concept}(a). 
The base fiber (SM980(4.5/80), Fibercore) has an initial diameter of 80~$\mu$m and is tapered down to 650~nm over a 1~mm length.
This geometry is specifically designed to facilitate trapping of Yb atoms in close proximity to the nanofiber surface using optical tweezers at 759~nm, the magic wavelength for the ${}^1\textrm{S}_0 \rightarrow {}^3\textrm{P}_0$ clock transition.
The tapered regions between the bare fiber and the nanofiber region are engineered to minimize radiation losses by ensuring adiabatic mode transformation.

Ultra-low-loss nanofibers at visible to near-infrared wavelengths have been routinely produced using an H${}_{2}$ and O${}_{2}$ flame.
Compared to other heating methods such as electric heaters or CO$_2$ lasers, this flame offers clean and uniform heating, enabling stable and reproducible fabrication of low-loss nanofibers. 
For instance, a transmission loss of 0.07 \% at 852~nm has been demonstrated using this approach~\cite{Horikawa2024-ra}. 
However, the combustion of H$_2$ with O$_2$ around optical fibers implants OH bonds into silica, leading to a significant transmission loss around 1380~nm as illustrated in Fig. \ref{fig:concept}(b).
This loss arises from the 1st overtone of the fundamental vibrational mode of Si-OH at 2760~nm~\cite{Humbach1996-pb}, and it poses a critical obstacle for low-loss nanofiber fabrication in the telecom band, especially at 1389 nm relevant to Yb transitions.

To resolve this issue while keeping the excellent controllability of the flame-based pulling, we replace the H${}_{2}$ gas with a D${}_{2}$ gas in the gas mixture. 
As an isotope of hydrogen, deuterium reacts similarly with oxygen to produce a high-temperature, clean flame suitable for nanofiber fabrication.
Although the D$_2$-O$_2$ flame similarly introduces OD bonds into the silica, the corresponding absorption band of Si-OD isotopically shifts from 1380~nm to 1860~nm as depicted in Fig.~\ref{fig:concept}(b).
This also leaves a transparent window between 1260~nm and 1660~nm free from overtone or combination tone absorption features of Si-OD~\cite{Stone1987-vf}, which makes it suitable for the other two Yb telecom-band transition wavelengths at 1480~nm and 1539~nm.

We experimentally confirm the effect of replacing the H$_2$-O$_2$ flame with a D$_2$-O$_2$ flame by measuring the transmission loss of 1389~nm light during the pulling process as shown in Fig.~\ref{fig:nanofiber}.
As the pulling time increases, corresponding to a reduction in fiber diameter, the transmission loss under the H$_2$-O$_2$ flame rises near-monotonically, reaching 8\% at the target diameter of 650 nm. 
In contrast, with the D$_2$-O$_2$ flame, the loss remains nearly constant below 1\%, which clearly demonstrates the effectiveness of the D$_2$-O$_2$ flame in suppressing OH-induced absorption.
Here we note that the transmission loss of 1550~nm is recorded at the same time to ensure the success of the pulling process and it exhibits a constant loss below 1\% for the both cases.
While the sensitivity of this measurement is limited by intensity fluctuations during the fabrication process, a cavity structure enhances sensitivity to optical losses and enables more precise quantification of the remaining absorption and fabrication-induced imperfections as discussed in the next section.
\section{A low-loss nanofiber cavity for a telecom-band wavelength}
\label{sec:cavity}

\begin{figure}[t]
    \includegraphics{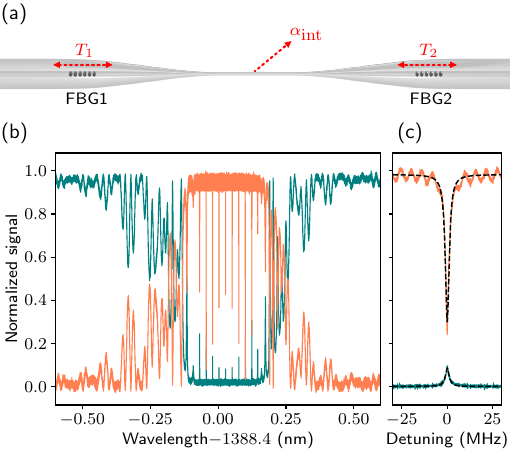}
    \centering
    \caption{
            (a)~The photon loss channels in a nanofiber cavity. Photons leak through FBG1 and 2 with transmittance $T_{1,2}$ and are scattered into free space or absorbed by silica inside the cavity with a decay rate $\alpha_\textrm{int}$.
            (b)~Transmission (green) and reflection (orange) spectra of a nanofiber cavity around 1389~nm. The free spectral range is 3.8~GHz, corresponding to the cavity length of 27~mm after the pulling.
            (c)~Zoomed-in spectra of a cavity resonance around the center region of the FBG stopband. The fringe pattern on the background of the reflection spectrum is due to the etalon effect in the measurement setup. The full width at half maximum of $2.9(2)$~MHz evaluated by fits with a Lorentzian function (black dashed lines) yields the finesse of the cavity to be $\mathcal{F}_\textrm{tot}=1.3(1)\times 10^3$. 
            }
    \label{fig:cavity}
\end{figure}

Having demonstrated the effectiveness of  the D$_2$-O$_2$ flame in suppressing the OH absorption, we now fabricate and characterize a nanofiber cavity at 1389~nm. 
The fabrication procedure follows the method described in~\cite{Ruddell2020-yn}. 
Briefly, two FBGs separated by 14~mm are first inscribed into the fiber core by exposing a deep ultraviolet laser through a phase mask. 
Each FBG is 8-mm long and has a stop band of 0.3~nm centered around 1389~nm, with a peak reflectivity of 99.9\%. 
Then, the section between the FBGs is formed into a nanofiber using the D$_2$-O$_2$ pulling technique described in the previous section.

To evaluate the round-trip loss of the nanofiber cavity, we adopt the method developed in~\cite{Horikawa2024-ra}, which we briefly describe here for completeness.
In a nanofiber cavity, photons are lost through three primary channels; transmission through the FBGs, denoted by $T_1$ and $T_2$, and intrinsic loss $\alpha_\mathrm{int}$, such as scattering into free space and material absorption, as illustrated in Fig.~\ref{fig:cavity}(a).
Thus, the total round-trip loss is given by  $\alpha_\textrm{tot}=T_1 + T_2 + \alpha_\textrm{int}$, which can be determined from the measured cavity finesse $\mathcal{F}_\textrm{tot}$ using the relation $\mathcal{F}_\textrm{tot}=2\pi/\alpha_\textrm{tot}$ in the regime $T_{1,2}\ll1$ and $\alpha_\textrm{int}\ll1$. 
%where the transmittance $T_{1,2}$ of each FBG and the intrinsic loss $\alpha_\textrm{int}$. 
%The total loss $\alpha_\textrm{tot}$ is obtained from the measured cavity finesse $\mathcal{F}_\textrm{tot}$ using the relation $\mathcal{F}_\textrm{tot}=2\pi/\alpha_\textrm{tot}$. 
To separate the contributions from each channel, we extract the FBG transmittances $T_{1}$ and $T_{2}$ from the corresponding on-resonant cavity reflectances $R_{1}$ and $R_{2}$, which represent coupling through FBG 1 and FBG 2, respectively, using the relation $R_{1,2}=(1-2T_{1,2}/\alpha_\textrm{tot})^2$. 
This enables us to isolate the intrinsic loss  $\alpha_\textrm{int}$, from which we define the intrinsic finesse as $\mathcal{F}_\textrm{int}=2\pi/\alpha_\textrm{int}$.
This figure of merit captures all losses except those due to mirror transmission, and serves as a critical metric for evaluating fabrication quality and material absorption.

Figures~\ref{fig:cavity}(b) and (c) show the transmission and reflection spectra of a nanofiber cavity around 1389~nm. 
The finesse is measured to be $\mathcal{F}_\textrm{tot}=1.3(1)\times 10^3$, corresponding to the total round-trip loss $\alpha_\textrm{tot}=0.48(3)\%$. 
We evaluate the transmittance of the FBGs by measuring reflection spectra from both ends of the cavity, from which we extract the intrinsic loss of $\alpha_\textrm{int}=0.31(2)\%$ and the intrinsic finesse $\mathcal{F}_\textrm{int}=2.0(1)\times 10^3$. 
When interfaced with Yb atoms, this cavity yields a projected cooperativity of 90, which is sufficient for high-fidelity and high-rate atom-photon entanglement generation \cite{Sunami2025-on}. %\textcolor{cyan}{[AG: I changed this sentence; maybe too specific? what do you think]}

\section{Conclusion}
\label{sec:conclusion}
We demonstrate the low-loss nanofiber fabrication at telecom-band wavelengths using a D$_2$-O$_2$ flame, which requires only minimal modifications to the conventional fabrication system using an H$_2$-O$_2$ flame. 
Building on this, we fabricate a nanofiber cavity at 1389~nm using the D$_2$-O$_2$ pulling method and achieve the intrinsic finesse of $\mathcal{F}_\textrm{int}=2.0(1)\times 10^3$. 
This corresponds to a measured round-trip loss that is 0.17\% higher than that of a previously reported nanofiber cavity at 852~nm~\cite{Horikawa2024-ra}. 
We attribute this excess loss primarily to unoptimized fabrication parameters and ambient moisture in the air.
To address this, further improvements can be made by 
controlling environmental conditions---for instance, by enclosing the fabrication system in dry air to reduce humidity-induced losses. 
Despite these limitations, the fabricated cavity already achieves a projected cooperativity of 90 when interfaced with Yb atoms, which is sufficient to reach the strong coupling regime and makes it suitable for use as an optical interconnect.

Importantly, since 1389 nm lies closest to the Si-OH absorption peak among the telecom-band transitions of Yb, the demonstrated method should be applicable to the other wavelengths, i.e., 1480 nm and 1539 nm \cite{Covey2019-ol, Li2024-bu}. 
Looking ahead, we plan to store the fabricated cavity in a dedicated container developed in Ref.~\cite{Horikawa2024-ra} and integrate it into an ultra-high-vacuum chamber for single Yb atom array experiments. 
Together, these results establish a key enabling technology for long-distance quantum communication and distributed quantum computing based on neutral-atom qubits.

\section*{Declaration of Competing Interest}
T.A. and A.G. are co-founders and shareholders of Nanofiber Quantum Technologies, Inc.~(NanoQT). Authors affiliated with NanoQT are employees or interns at NanoQT at the time of their contributions.

\section*{acknowledgements}
We thank S.~Umezawa for the technical support on the puller system development, and N.~Shiraishi, S.~Okamura, H.~Iida, T.~Matsui and M.~Shimasaki for the assistance of the nanofiber cavity fabrication. This work was supported by JST Moonshot R\&D (Grant Number JPMJMS2268).

\bibliography{refs}

\end{document}